\documentclass{eas}
\usepackage{graphicx}

\begin{document}

\title{Long-term spectropolarimetric monitoring of the cool supergiant Betelgeuse} 
\runningtitle{Spectropolarimetric monitoring of Betelgeuse} 
\author{I. Bedecarrax}\address{IRAP, Universit\'e de Toulouse, CNRS, Toulouse, France}
\author{P. Petit}\address{IRAP, Universit\'e de Toulouse, CNRS, Toulouse, France}
\author{M. Auri\`ere}\address{IRAP, Universit\'e de Toulouse, CNRS, Tarbes, France}
\author{J. Grunhut}\address{ESO, Garching bei M\"unchen, Germany}
\author{G. Wade}\address{Department of Physics, Royal Military College of Canada, Kingston, Ontario, Canada}
\author{A. Chiavassa}\address{Laboratoire Lagrange, Universit\'e de Nice Sophia-Antipolis, CNRS, Observatoire de la C\^ote d'Azur, Nice, France}
\author{J.-F. Donati}\address{IRAP, Universit\'e de Toulouse, CNRS, Toulouse, France}
\author{R. Konstantinova-Antova}\address{Institute of Astronomy and NAO, Bulgarian Academy of Sciences, Sofia, Bulgaria}
\author{G. Perrin}\address{Observatoire de Paris, LESIA, Meudon, France}

\begin{abstract}
We report on a long-term monitoring of the cool supergiant Betelgeuse, using the NARVAL and ESPaDOnS high-resolution spectropolarimeters, respectively installed at Telescope Bernard Lyot (Pic du Midi Observatory, France) and at the Canada--France-Hawaii Telescope (Mauna Kea Observatory, Hawaii). The data set, constituted of circularly polarized (Stokes V) and intensity (Stokes I) spectra, was collected between 2010 and 2012. We investigate here the temporal evolution of magnetic field, convection and temperature at photospheric level, using simultaneous measurements of the longitudinal magnetic field component, the core emission of the Ca II infrared triplet, the line--depth ratio of selected photospheric lines and the radial velocity of the star.
\end{abstract}

\maketitle

\section{Introduction}

\begin{figure}
\centering
\includegraphics[width=10cm]{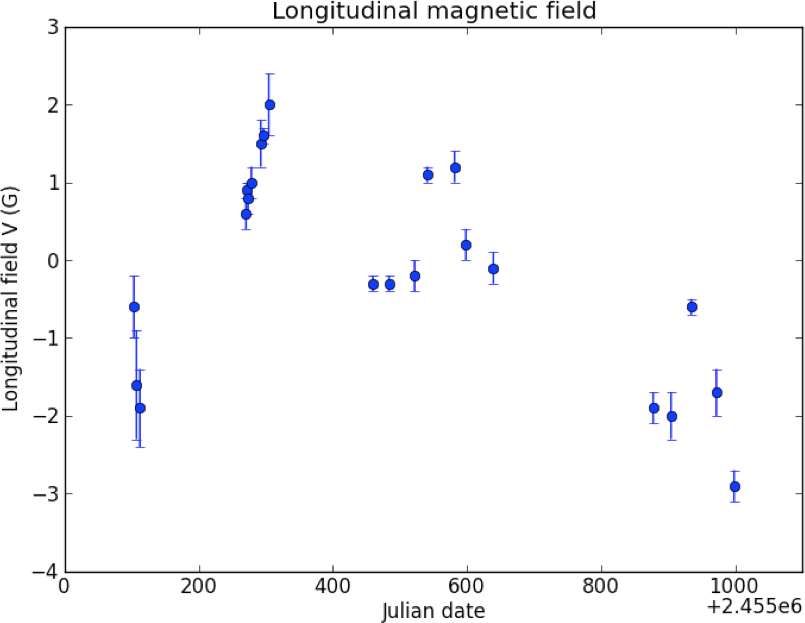}
\caption{Longitudinal magnetic field, as a function of the Julian date.}
\label{fig:blong}
\end{figure}

The cool supergiant Betelgeuse is a rare laboratory to test stellar dynamo models in an extreme region of their parameter space, for at least three reasons. First, its extremely slow rotation (with a rotation period of about 17~yr, \cite{uitenbroek98}) implies that Betelgeuse is probably experiencing very limited (if any) action of a solar-like dynamo, as opposed to a number of cool evolved stars with rotation periods as short as a few days (\cite{petit04a}, \cite{petit04b}, \cite{konstantinova08}). With a very advanced evolutionary status, Betelgeuse is also benefitting from a very large pressure scale height (compared to Sun-like stars), resulting in the presence of no more than a few giant and stable convection cells at photospheric level (\cite{schwarzschild75}, \cite{chiavassa10}). This second asset dramatically enhances the detectability of dynamo-related magnetic features with sizes comparable to the convective spatial scale, e.g. magnetic regions generated through a turbulent dynamo (\cite{vogler07}). Finally, the very large radius of Betelgeuse ensures that any remnant of a possible strong main-sequence magnetic field is likely negligible near the photosphere owing to a large dilution factor, as opposed to less evolved probable descendants of Ap/Bp magnetic stars of intermediate mass (\cite{auriere08}, \cite{auriere12}, \cite{tsvetkova13}). 

Betelgeuse is therefore an ideal target to unveil the genuine action of a turbulent dynamo and test its possible role in the sustained mass loss of cool supergiants. We present here the results of a long-term monitoring of the surface magnetic field of Betelgeuse.

\section{Observations and data reduction}

\begin{figure}
\centering
\includegraphics[width=6.1cm]{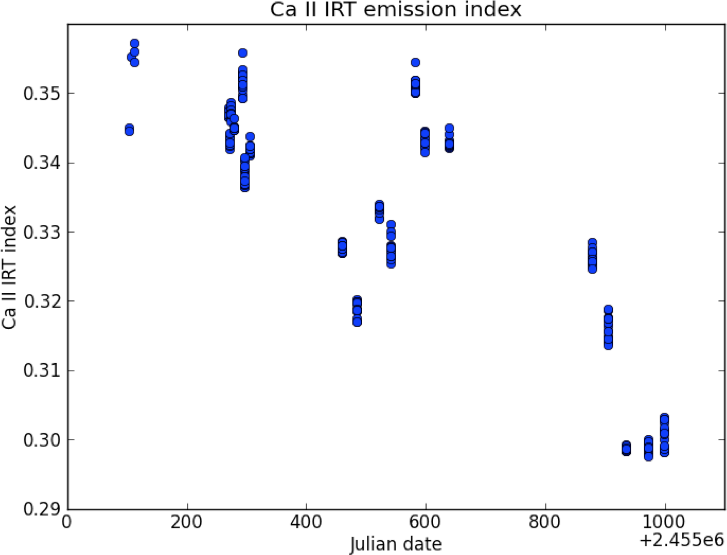}
\includegraphics[width=6.1cm]{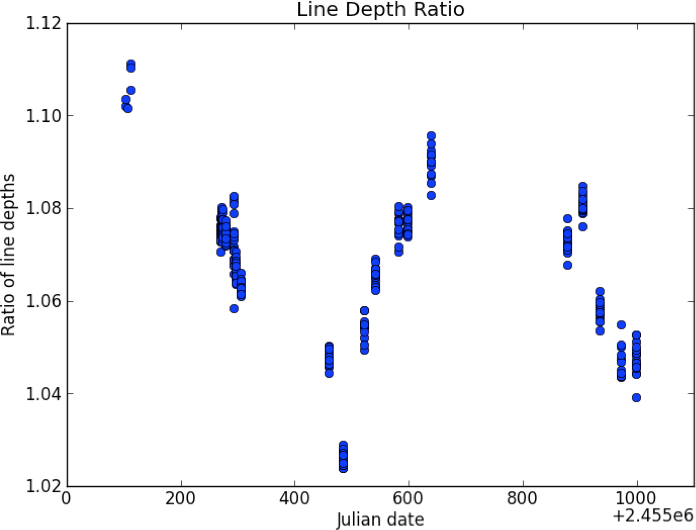}
\includegraphics[width=6.1cm]{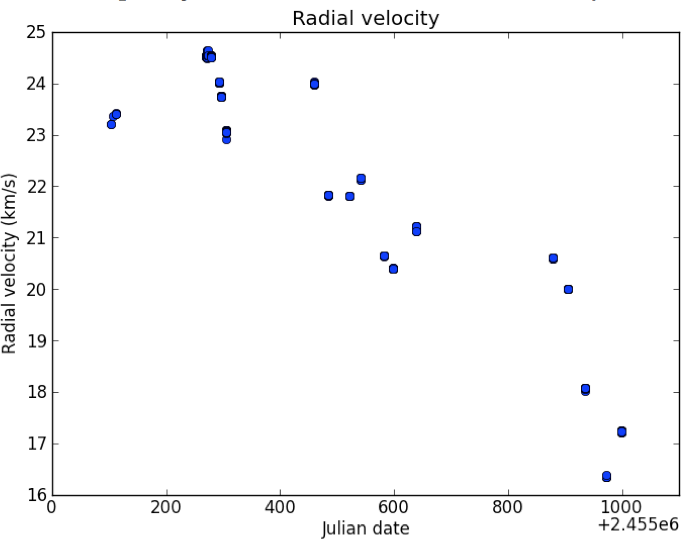}
\includegraphics[width=6.1cm]{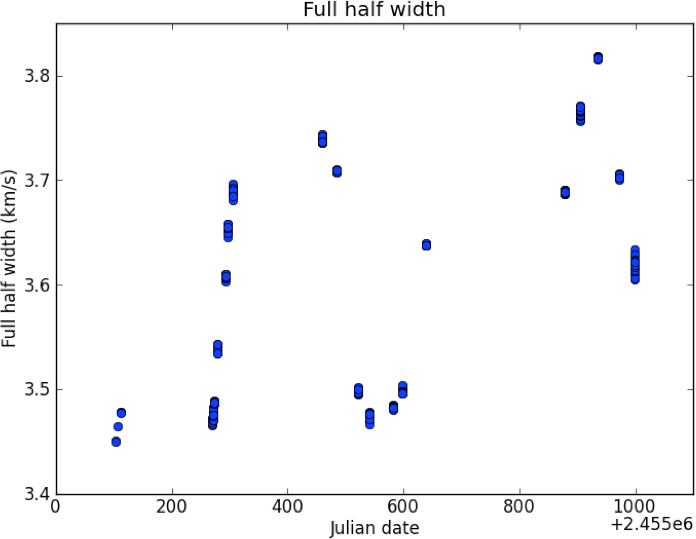}
\caption{Core emission of the Ca\,{\sc ii} infrared triplet (top left), line depth ratio of V\,{\sc i}~625.183 nm over Fe\,{\sc i}~625.257 nm (top right), radial velocity (bottom left) and FWHM of Stokes I LSD profiles (bottom right). All quantities are plotted as a function of the Julian date.}
\label{fig:tracers}
\end{figure}

The observations were gathered with the NARVAL and ESPaDOnS high-resolution spectropolarimeters. These twin instruments cover the whole optical domain (370 nm - 1,000 nm) in one exposure, with a spectral resolution of $\approx65,000$. Unpolarized (Stokes I) and circularly-polarized (Stokes V) spectra were simultaneously recorded. 29 spectra were obtained with ESPaDOnS, from 2009 Sep. 28 to 2012 Feb 13.  284 spectra were collected with NARVAL, from 2010 Mar 14 to 2012 Mar 11.

To reach the S/N required to detect the weak magnetic field of Betelgeuse, daily observations were averaged together and processed by the Least-Square Deconvolution method (LSD hereafter, \cite{donati97}). Using a list of spectral lines roughly matching a stellar photospheric model for the effective temperature of Betelgeuse, we calculated averaged line profiles in Stokes I and V. Thanks to this cross-correlation method, we detect polarized Zeeman signatures in most observations, with a typical amplitude of $3 \times 10^{-5}$ of the continuum (\cite{auriere10}). 

\section{Longitudinal magnetic field}

We use the Stokes V LSD profiles to measure the longitudinal component of the photospheric magnetic field, using the centre-of-gravity technique described by \cite{rees79}. A magnetic field is detected in most available observations, with values ranging from -3 G to +2 G (Fig. \ref{fig:blong}). Temporal changes in the field strength occur over timescales as short as a few weeks (see e.g. the steep increase observed during the spring of 2010). Early observations obtained in 2009 unveil a negative field polarity, soon replaced by a positive polarity in 2010. Measurements from 2011 are also mostly consistent with a positive field, and a negative polarity is observed again in 2012, highlighting the overall decrease of the field strength from 2010 to 2012.

\section{Core emission of the Ca\,{\sc ii} infrared triplet }

We estimate the core emission in the  Ca\,{\sc ii} IRT by computing an emission index similar to the one proposed by \cite{petit13}. Steep fluctuations of the index are recorded over a few weeks, with a global decrease from 2009 to 2012 (Fig. \ref{fig:tracers}, top left). The observed variations are correlated with the H$\alpha$ core emission (not shown here). No obvious correlation is observed with temporal variations of the longitudinal magnetic field strength.

\section{Line depth ratio}

Following \cite{gray08}, we measured the ratio of line depths of V\,{\sc i}~625.183 nm and Fe\,{\sc i}~625.257 nm, taken as a surface temperature proxy. The temporal evolution of the ratio is smooth and correlated with the core emission of the Ca\,{\sc ii} IRT (Fig. \ref{fig:tracers}, top right). The range and timescale of the observed fluctuations are consistent with previous findings by Gray.  

\section{Radial velocity and FWHM of Stokes I LSD profiles}

Radial velocities and FWHM of photospheric lines were estimated by simply adjusting a gaussian function on Stokes I LSD profiles. If radial velocities vary within typically one month, longer-term variations are observed as a global decrease from 2010 to 2012, from 24 km/s to 17 km/s (Fig. \ref{fig:tracers}, bottom left). Fast fluctuations are also observed in the FWHM, but longer trends are less obvious in this proxy, with an increase of $\approx0.4$ km/s over 2 years (Fig. \ref{fig:tracers}, bottom right).

\end{document}